\documentclass{article}

\usepackage{graphicx}
\usepackage{fullpage,amsmath,amssymb,latexsym}
\usepackage{multirow}
\usepackage{natbib}
\begin{document}
\title{Empirical Models of Pressure and Density in Saturn's Interior: Implications for the Helium Concentration, its Depth Dependence, and Saturn's Precession Rate}
\author{Ravit Helled$^{1,**}$, Gerald Schubert$^1$, and John D. Anderson$^2$\\
\small{$^1$Department of Earth and Space Sciences and Institute of Geophysics and Planetary Physics,}\\
\small{University of California, Los Angeles, CA 90095Ð1567, USA}\\
\small{$^2$Jet Propulsion Laboratory } \\
\small{California Institute of Technology, Pasadena, CA 91109}\\
\small{$^{**}${\it corresponding author, E-mail address: rhelled@ess.ucla.edu}}\\
}

\date{}
\maketitle 



\begin{abstract}
We present 'empirical' models (pressure vs. density) of Saturn's interior constrained by the gravitational
coefficients $J_2$, $J_4$, and $J_6$ for different assumed rotation
rates of the planet. The empirical pressure-density profile is interpreted in terms of a hydrogen and helium physical equation of state to deduce the hydrogen to helium ratio in Saturn and to constrain the depth dependence of helium and heavy element abundances. 
The planet's internal structure (pressure vs. density) and composition are found to be insensitive to the assumed rotation rate for periods between 10h:32m:35s and 10h:41m:35s. We find that helium is depleted in the upper envelope, while in the high pressure region (P $\gtrsim$ 1 Mbar) either the helium abundance or the concentration of heavier elements is significantly enhanced. 
Taking the ratio of hydrogen to helium in Saturn to be solar, we find that the maximum mass of heavy elements in Saturn's interior ranges from $\sim$ 6 to 20 M$_{\oplus}$.\\
The empirical models of Saturn's interior yield a moment of inertia factor
varying from 0.22271 to 0.22599 for rotation periods
between 10h:32m:35s and 10h:41m:35s, respectively. A long-term precession rate of about 0.754'' yr$^{-1}$ is found to be consistent with the derived moment of inertia values and assumed rotation rates over the entire range of investigated rotation rates. This suggests that the long-term precession period of Saturn is somewhat shorter than the generally assumed value of 1.77$\times 10^6$ years inferred from modeling and observations.

\end{abstract}

\newpage
\section{Introduction}
Models of Saturn's interior based on pre-Cassini values of the planet's gravitational coefficients and the equation of state of \citeauthor{Saumon95} (1995) have been presented by several authors (e.g., \citealp{Fortney03,Saumon04}). Nevertheless, there is still considerable uncertainty in Saturn's internal structure due to incompleteness and lack of precision in our knowledge of Saturn's gravitational field, the absence of information on Saturn's abundance of helium and heavy elements, and the uncertainty in the equation of state of hydrogen-helium mixtures at high pressures and temperatures. For example, the size and mass of Saturn's heavy element core and the depth of the transition from molecular to metallic hydrogen are unknown. An additional source of uncertainty in determining Saturn's internal structure has arisen with the realization that we do not know the rotation rate of the deep interior (\citealp{Gurnett07}).\\
In this paper we seek to provide reference radial profiles of pressure ($p$) and density ($\rho$) in Saturn's interior based on mass, radius, gravitational coefficients from the analysis of Cassini (\citealp{Jacobson06}) and other data. These interior models are non-unique because of Saturn's unknown rotation rate and uncertainties in the gravitational coefficients, but as we will show, they provide tight constraints on the real distribution of pressure and density inside Saturn, on properties of Saturn such as its moment of inertia and precession rate, and on the composition of the planet.\\ 
Because we represent the radial profile of density in Saturn as a polynomial function of the radial coordinate $s$, the mean radius in the interior, the profile is independent of uncertainties and assumptions about the equation of state of hydrogen-helium mixtures. The same is true about the radial profile of pressure which follows from integration of the hydrostatic equation using the radial density profile. Elimination of $s$ between $\rho(s)$ and $p(s)$ gives an 'empirical' equation of state (EOS), $\rho=\rho(p)$. The empirical EOS is dependent only on the assumed internal rotation rate of Saturn and it is uncertain only by the truncation of the gravitational field representation and the errors in the gravitational coefficients. However, the representation of the radial profile of density by a 6'th degree polynomial function of mean radius may be inadequate to account for all the features of Saturn's actual interior such as a density discontinuity at the surface of a heavy element core. With the exception of the small effect of the unknown Saturnian rotation rate, the empirical EOS we derive provides a unique polynomial function model of the radial distribution of pressure and density inside Saturn.\\
In the next section interior models ($\rho(s)$, $p(s)$, $\rho(p)$) of Saturn are derived using the 'theory of figures' (Zharkov \& Trubitsyn, 1978). The interior models fit the atmospheric model of Lodders \& Fegley (1998) and the Saturnian gravitational moments. The models make no assumptions about the planet's composition or its radial dependence. Since the rotation rate of Saturn's deep interior is still unknown we follow \citeauthor{And07} (2007) and assume rotation periods between 10h:32m:35s and 10h:41m:35s.  In section 3, the physical equation of state of \citeauthor{Saumon95} (1995) is used to infer the hydrogen to helium ratio of the planet and its dependence on radius based on comparison with the empirical EOS of this paper. Section 4 presents the values of moment of inertia and the precession period of Saturn's rotation axis  predicted by the empirical EOS for the range of rotation periods studied. We conclude with a general discussion of the results.

\section{Interior Models: Finding Radial Profiles of Density and Pressure in Saturn's Interior -  An Empirical Equation of State}
The calculation of the interior models proceeds in three steps. First, the measured gravitational field and polar radius are used to obtain the reference geoid or the effective gravitational potential function $U$, where
\begin{eqnarray}
U &=& V + Q \nonumber \\
Q &=& \frac{1}{2} \omega^2 r^2 \sin^2 \theta \nonumber \\
V &=& \frac{G M}{r} \left( 1 - \sum_{n=1}^\infty \left( \frac{a}{r} \right)^{2 n} J_{2 n} P_{2 n} \left( \cos \theta  \right)  \right).
\label{U}
\end{eqnarray}
In (1), $V$ is the gravitational potential, $Q$ is the centrifugal potential, and $\omega$ is the angular velocity of rotation. We take the rotation to be that of a solid body with constant angular velocity. Though the atmosphere is differentially rotating, our models remain relevant as long as the differential rotation is shallow and does not affect the deep interior. Further in (1), (r, $\theta, \phi$) are spherical polar coordinates, $G$ is the gravitational constant and $M$ is the total planetary mass. For a rotating fluid in hydrostatic equilibrium, only the even zonal harmonics are stimulated, and the gravitational potential $V$ can be represented as an expansion in even Legendre polynomials $P_{2 n}$ (\citealp{Kaula1968,ZT1978}). For pure rotation, the longitudinal angle $\phi$ does not enter in $U$. The constants that define $V$ for a particular planet are the gravitational constant times the total planetary mass $G M$, an equatorial radius $a$, and the harmonic coefficients $J_{2 n}$, which can be inferred from Doppler tracking data of a spacecraft in the planet's vicinity, such as the Cassini orbiter of Saturn.   

The reference geoid is defined as the surface of constant effective potential $U$ with occultation polar radius of 54,438 km. In the absence of any published Cassini data on occultation radii, we used the published Voyager value for the polar radius of the 100 mbar isosurface, along with its standard error (Lindal et al., 1985; \citealp{Nich95}). The polar radius is expected to be relatively independent of rotation rate and atmospheric winds, and is held fixed in our models for the reference geoids. For each rotation period there is a different surface shape of equal gravitational potential in which the equatorial radius changes with respect to the rotation period. Occultation data can define the equatorial radius of the 100 mbar isosurface only for an assumed rotation period of Saturn. As shown by Anderson \& Schubert (2007), the altitude above the reference geoid of the measured 100 mbar surface can vary by hundreds of kilometers at the equator, depending on the assumed rotation period (Lindal et al., 1985; \citealp{Hub97}).
The mean radius $R$ for the 100 mbar isosurface is defined by the radius of a sphere that has the same mean density $\rho_0$ as Saturn. The mean density is dependent on the rotation rate as shown in Table 1. For a more rapidly rotating planet, the equatorial bulge is greater and the mean density is lower for the same total mass. \\
{\bf [Table. 1]}\\
From the Cassini mission, the $GM$ for Saturn is 37931208 km$^3$s$^{-2}$ (Jacobson et al., 2006), and with a modern value for the gravitational constant $G$ of $6.674215\times10^{-11}$ m$^3$s$^{-2}$kg$^{-1}$ (\citealp{Gundlach00}), the total mass of Saturn $M$ is $5.683246\times10^{26}$ kg, with a fractional uncertainty in $G$ equal to 14 ppm. The mean density $\rho_0$ is defined as this total mass divided by the volume of the fifth order reference geoid. The smallness parameter $m$ follows from its definition $m=\omega^2R^3/GM$ (Zharkov \& Trubitsyn, 1978), where $\omega$ is the angular velocity associated with the periods of rotation in Table 1. The similar smallness parameter $q$, given by $\omega^2 a^3/GM$, is used for the calculation of the reference geoid. Self-consistent values of $a$ and $q$ are found by iterating the calculation of the reference geoid as discussed in the caption of Table 1. The gravitational coefficients $J_2$, $J_4$ and $J_6$ are simultaneously determined from the iterations using the values of the observed gravitational coefficients $J_2$, $J_4$, and $J_6$ for the reference equatorial radius of 60,330 km (Jacobson et al., 2006) according to the procedure discussed in the caption of Table 1. Values of $J_8$ and $J_{10}$ for the fifth order calculation are obtained by extrapolation from the lower-degree coefficients. 

The characteristic pressure in the interior is defined by $p_0$=GM$\rho_0/R$ (Zharkov \& Trubitsyn, 1978). Given the parameters of Table 1, the normalized mean radius $\beta$ is defined by $s/R$, where $s$ is the mean radius in the interior, the normalized mean density $\eta(\beta)$ is $\rho(s)/\rho_0$, and the normalized pressure $\xi(\beta)$ is $p(s)/p_0$. From the theory of figures (Zharkov \& Trubitsyn, 1978), a particular interior model is characterized by an assumed density distribution $\eta(\beta)$ and smallness parameter $m$, and by the shape of level surfaces over the interval $0\le\beta\le1$. The basic idea of the interior calculations is to start with a best guess for $\eta(\beta)$, compute the level surfaces in the interior, and then evaluate the harmonic coefficients $J_2$, $J_4$, and $J_6$ at the surface for $\beta$ equal to unity. The differences between the calculated coefficients and the observed surface values from Table 1 are used to correct the density function, and the process is iterated to convergence. Further discussion of the density distribution $\eta(\beta)$ and the interior model is given below. 

Once a density distribution that matches the observed gravitational coefficients is available, the pressure in the interior is obtained by integration of the equations of hydrostatic equilibrium and mass continuity. The equations to the first order in $m$ are (Zharkov \& Trubitsyn, 1978),
\begin{equation}
\frac{1}{\eta} \frac{d \xi}{d \beta} = - \frac{\alpha}{\beta^2} + \frac{2}{3} m \beta
\label{etaODE}
\end{equation}
\begin{equation}
\frac{d \alpha}{d \beta} = 3 \eta \beta^2,
\label{alphaODE}
\end{equation}
where the parameter $\alpha$ is the normalized mass $M(\beta)/M$ internal to a level surface labeled by the mean fractional radius $\beta$. The normalized axial moment of inertia for the planet $\gamma$ is also available from the density distribution by the integration
\begin{equation}
\gamma = \frac{C}{MR^2} = 2\int_0^1\eta \beta^4 d\beta
\label{MOI}
\end{equation}
where $C$ is the axial moment of inertia. Because the internal density distribution matches the observed gravitational coefficients, the normalized moment of inertia from Eq.~4 
is consistent with those coefficients, and with the rotation period. 

\subsection*{Atmospheric boundary condition}

We represent the internal density distribution by a single sixth degree polynomial with the first degree term missing. 
Such a polynomial contains as many unknowns as the degrees of freedom imposed by the atmospheric density profile and its connection to the interior. The density distribution near the surface is based on two degrees of freedom. The data for the interior consist of the three gravitational harmonics J$_2$, J$_4$, J$_6$ and the measured mass and mean radius of Saturn. The rotation rate is a free parameter, and with an assumed density distribution as a function of radius, there are three observational constraints on the polynomial. The implementation of the method of level surfaces finds a one-to-one match between the polynomial and the three gravitational harmonics at a given rotation rate. The sixth degree polynomial, with the first degree term set to zero, has six coefficients that can be fit to the five measurements (three harmonics plus two atmospheric constraints) by the method of nonlinear least squares. The indeterminacy is eliminated by imposing a condition on the polynomial that all the measured mass is included between the center and the surface ($0 \leq \beta \leq 1$). Hence there are five measurements and five free parameters in the fitting model. Each of our models for a given rotation rate is unique. For polynomials of degree $n>6$ there are $n-6$ degrees of indeterminacy in the fitting process. A steep increase of density at the center could be imposed as a boundary condition on a higher-degree polynomial, but the resulting best-fit model would then depend on that assumed boundary condition. With the sixth degree polynomial, an interior model that fits all the available data and that is also free of any physical constraints on the interior can be obtained.
The derivative of the density goes to zero at the center, and it must be negative everywhere else on the interval $0<\beta\le1$. Otherwise the density would decrease with increasing depth, which is a physical impossibility. We further eliminate the sixth degree polynomial coefficient by insisting that all the mass be used up during the integration of equation (3) 
from the center to the surface. The resulting polynomial with five free coefficients is,
\begin{eqnarray}
\eta &=& 3 \beta^6 + k_0 \left( 1 - 3 \beta^6 \right) + k_2 \beta^2 \left( 1 - \frac{9}{5} \beta^4 \right) + k_3 \beta^3 \left( 1 - \frac{3}{2} \beta^3 \right) + \nonumber \\
&&k_4 \beta^4 \left( 1 - \frac{9}{7} \beta^2  \right) + k_5 \beta^5 \left( 1 - \frac{9}{8} \beta   \right).
\label{poly}
\end{eqnarray}
During the calculation of the level surfaces, we check to make sure that the derivative of the polynomial is everywhere negative, but that is not imposed as an additional constraint. If the derivative were positive anywhere on the interval, we would take that as a proof that the single polynomial is an inappropriate approximation to the true density distribution in the interior.

In previous models (Anderson \& Schubert, 2007) the polynomial and its derivative were set to zero at the surface. For purposes of better matching the interior polynomial to the atmosphere, we use the model atmosphere in Table 9.2 of Lodders \& Fegley (1998), and derive least-squares normal equations for the polynomial coefficients. We interpolate in the table and derive the normalized density at four values of $\beta$ (0.9985, 0.9990, 0.9995, 1.0), where the value 1.0 represents the 100 mbar level. For the mean radius and mean density associated with the 10h:32m:35s rotation period (Table 1), the corresponding atmospheric densities $\eta$ are (0.0002666, 0.0001719, 0.0000945, 0.0000408). These four points are fitted by least squares in combination with the normal equations associated with the observed gravitational coefficients $J_2$, $J_4$, $J_6$. The errors on the atmospheric densities are taken at 100\% and the errors on the gravitational coefficients are given by their converged covariance matrix from the fits to Cassini and other data (Jacobson et al., 2006).

The partial derivatives of the atmospheric density with respect to the polynomial coefficients are simply the coefficients in the polynomial of equation (5). 
These partial derivatives are collected into a $4\times5$ matrix $A$, where each row of the matrix corresponds to a particular observed atmospheric density. The atmospheric least-squares problem is linear, and the polynomial coefficients that correspond to the model atmosphere can be obtained at once. However, the rank of the $A$ matrix is not four, but two. The atmosphere imposes two constraints on the polynomial, similar to the simpler but less satisfactory assumption that both the density and its derivative are zero at the surface. The polynomial coefficients from the atmosphere alone can be obtained by singular-value decomposition of the matrix $A$ and the computation of its pseudo inverse (\citealp{Lawson74}), but the resulting polynomial is not useful for extrapolation to values of $\beta$ less than 0.9985. The three observed gravitational coefficients must also be introduced for a meaningful determination of the polynomial. The atmospheric model serves only as a boundary condition on the deeper interior model of interest.

We form a diagonal weighting matrix $W$ with the inverse squares of the four $\eta$ data on the diagonal (100\% error). The atmospheric normal equations can then be written as (\citealp{Lawson74}),
\begin{equation}
(A^TWA)x = A^TWz,
\label{normatmos}
\end{equation}
where the superscript $T$ indicates a transpose, $x$ is a $5\times1$ column matrix containing corrections to the assumed polynomial coefficients, and z is a $4 \times 1$ column matrix containing the corresponding residuals to the atmospheric values of $\eta$. These atmospheric normal equations are combined with the normal equations for the gravitational coefficients, where the gravitational normal equations are designated by a subscript J. The best estimate $x$ of the corrections to the polynomial coefficients from the two data sets is,
\begin{equation}
(A^TWA + A_J^TW_JA_J)x = A^TWz + A_J^TW_Jz_J
\label{normcomb}
\end{equation}
The gravitational $A_J$ matrix is a $3 \times 5$ matrix containing partial derivatives of the three gravitational coefficients with respect to the five polynomial coefficients. The weighting matrix $W_J$ is the inverse of the $3 \times 3$ covariance matrix from the data fits for the gravitational field (Jacobson et al., 2006). The $3 \times 1$ matrix $z_J$ contains the residuals for the observed $J_2$, $J_4$, $J_6$ corresponding to the current estimate of the polynomial. The process is iterated until it converges. The inverse of the converged matrix $(A^TWA + A_J^TW_JA_J)$ is the covariance matrix for the five polynomial coefficients. It can be mapped onto the polynomial for purposes of obtaining error bars on the density distribution in the interior. The fractional error is largest near the surface where the assumed error is 100\%, but where the density is small. \\
{\bf [Fig. 1]}\\
Figure 1 shows the uncertainty in the $\rho(p)$ relation for a rotation period of 10h:32m:35s. The solid line presents the computed $\rho(p)$ relation, and the dotted and dashed-dotted curves present the $\rho(p)$ relation when the error (uncertainty in the EOS) is added and subtracted, respectively. The area between these two curves represents the empirical EOS derived from the interior model. As can be seen from the figure, the difference is largest in the lowest pressure region. This is due to the 100\% error assumed on the atmospheric density. As the pressure increases the error decreases and the three curves overlap. A smaller error in the atmospheric model would lead to a more accurate $\rho(p)$ relation in the low pressure regions. Because the residuals for the gravitational harmonics are much smaller than their standard errors, the polynomial in the deep interior is insensitive to the relative weighting of the gravitational and atmospheric data.  Even in the outer atmospheric layers, a 100\% error in the atmospheric density produces only a Log 2 deviation from the best-fit polynomial. We conclude that the empirical density function is robust in the high pressure region and more uncertain in the upper part of Saturn's envelope. The covariance matrix also can be mapped onto the pressure distribution by the integration of equation (2) 
and the random error in the empirical equation of state can be estimated for a fixed value of the period. The systematic error caused by uncertainties in the rotation period of about plus six and minus one minute about our preferred period of 10h:32m:35s (Anderson \& Schubert, 2007) is given by the spread in parameters across Table 1 and Table 2. However, the empirical equation of state is not particularly sensitive to this relatively large systematic error in the period, let alone the random error. Its determination is robust. 

\subsection*{Calculation of level surfaces and the gravitational harmonics}

With a given value of the smallness parameter $m$ and the assumed density distribution $\eta(\beta)$, the level surfaces for constant internal potential can be evaluated and the surface harmonics can be computed from a series approximation in $m$ to the equation (Zharkov \& Trubitsyn, 1978),
\begin{equation}
Ma^nJ_n = - \int_\tau \rho(r) r^n P_n(\cos \theta) d \tau,
\label{Jn}
\end{equation}
where the integration is carried out over the volume $\tau$, $r$ is the radius, $\theta$ is the polar angle or colatitude, and $P_n$ is the Legendre polynomial of degree $n$.\\
We have coded the level surface theory, which is given by Zharkov \& Trubitsyn (1978) to the fifth order, but we truncate to terms in third order. From an assumed density distribution $\eta(\beta)$ (equation 5) the gravitational harmonics $J_2$, $J_4$, and $J_6$ are evaluated and the three residuals for the $z_J$ matrix become available. The $A_J$ matrix is obtained by finite differencing the polynomial coefficients one at a time and by running the level surface code five times to obtain an estimate of the partial derivatives. The covariance matrix $W_J^{-1}$ for the observed gravitational harmonics in units of 10$^{-6}$ is given by (Jacobson et al., 2006),
\begin{equation}
W_J^{-1} = \bordermatrix{& J_2 & J_4 & J_6 \cr
                      J_2 & 0.0746 & 0.6166 & 1.3618 \cr
                      J_4 & 0.6166 & 7.6984 & 23.2222 \cr
                      J_6 & 1.3618 & 23.2222 & 93.0205 \cr }
\end{equation}

{\bf [Table. 2]}\\
By iterating with equation (7) the polynomial coefficients of Table 2 are obtained as a best fit to the gravitational harmonics and the atmospheric data. The converged residuals for the harmonic coefficients also are given in Table 2. The fit is not perfect because there is a trade-off between the fit to the atmosphere and the fit to the harmonics. However, the fact that the two data sets are satisfied well within their respective standard errors lends credibility to the interior density distribution.\\
The density distribution is found to be relatively insensitive to the assumed rotation rate for periods between 10h:32m:35s and 10h:41m:35s. 
Figure 2 presents the pressure-density relation from the interior models. We present the results for rotation periods of 10h:32m:35s and 10h:41m:35s, the shortest and longest rotation periods considered. The density functions are very much alike even with a nine minute difference in rotation period. To see more clearly the difference in the two functions, Figure 2 divides the pressure range into four regions. If the density-pressure relations were presented for the entire pressure range of the planet, the two curves would overlap. \\
{\bf [Fig. 2]}\\
The values obtained for the normalized axial moment of inertia (see equation (4)) in each of the interior models are also presented in Table 2. The moment of inertia of the Saturn models depends weakly on the rotation rate for the range of periods considered.\\

\section{Hydrogen to Helium ratio in Saturn}
 
Saturn is mostly a convective mixture of hydrogen and helium with a minor amount of heavy elements (\citealp{Hubbard68}; \citealp{Guillot94}). 
In this section we use the empirical EOS (pressure-density profile) derived from the interior models to investigate the hydrogen-helium distribution in Saturn's interior. Using the EOS of Saumon et al. (1995), we look for the hydrogen-helium mixing ratio that produces a pressure-density profile similar to the one found by the interior models. 
Theoretical models of Saturn's interior suggest that the planet contains $\sim$10 - 30 M$_\oplus$ of heavy elements (\citealp{Saumon04}), however, the exact amount of the material and its composition are unknown. For simplicity, when fitting the empirical pressure-density models we include only hydrogen and helium. In that case, the high-Z material is manifest as a higher helium mass fraction (\citealp{Guillot07}; \citealp{Baraffe08}), so an increase of the helium mass fraction above the proto-Sun value can reveal the amount of high-Z material in the interior. \\
To produce a physical EOS that can be compared to the p-$\rho$ relation from our interior models we take an adiabatic EOS of a homogeneous  hydrogen-helium mixture. An adiabatic EOS is justified since Saturn's interior is expected to be convective, with an opacity increase with increasing pressure and temperature (\citealp{Hubbard73}). Although convection might be suppressed by compositional gradients, condensation or rotation, Saturn interior models predict either no radiative zone or a very narrow one (Guillot et al.~, 2004). \\
An adiabat can be produced once the value of the entropy is known. The entropy of a hydrogen-helium mixture is given by (\citealp{Saumon95}), 
\begin{equation} 
S(p,T,X)= XS_H(p,T)+YS_{He}(p,T)+S_{mix}(p,T,X)
\end{equation} 
where $X$ is the mass fraction (mass mixing ratio, mass of atoms over the total mass) of hydrogen, $S_H$ is the entropy of hydrogen, $Y\equiv1-X$ and $S_{He}$ are the mass fraction and entropy of helium, respectively, and $S_{mix}$ is the entropy of the mixture. The entropy of the mixture is determined using the constraint on the temperature at the 1 bar level in Saturn's atmosphere.
The commonly used value for the Saturnian temperature at the 1 bar pressure level is 134.4 K (\citealp{Lindal92}), but it has been suggested by \citeauthor{Guillot99} (1999) that this temperature could be as high as 145 K. 
In addition, the temperature at the pressure of 1 bar is obtained from radio-occultation measurements for an assumed helium to hydrogen ratio. Temperature is not directly observed but inferred based on an assumption about the composition of the planet's atmosphere. Because of the uncertainty in the surface temperature, we follow \citeauthor{Saumon04} (2004), and take Saturn's temperature at 1 bar to range from 130 K to 145 K. The temperature at 1 bar is significant because it determines the entropy, $S(p,T,X)$, and therefore the adiabat. Once the entropy is determined, we obtain a density function $\rho(p,X,S)$ that can be compared to the empirical density function derived in the previous section. 

We compute the adiabats over Saturn's interior pressure range for hydrogen to helium mixing ratios ranging from pure hydrogen to pure helium.
We determine the composition which gives the best fit, in the least-squares sense, of the hydrogen-helium mixture EOS to the empirical p-$\rho$ relation found by the interior models (section 2). We find that the minimum of the standard deviation function is broad and as a result we also report the range in composition over which the best fit value changes by $\pm10 \%$ (uncertainty factor). This range also accounts for the uncertainty in the empirical EOS.\\
{\bf [Fig. 3]}\\
Figure 3 shows the difference between the empirical EOS and the computed density-pressure relations which give the best fit for rotation periods between 10h:32m:35s and 10h:41m:35s, and different temperatures at the 1 bar level.  The dotted, solid, dashed-dot and dashed curves represent temperatures of 130, 135, 140 and 145 K, respectively. 
The larger density difference between the empirical and physical p-$\rho$ relations in the deep interior suggests that a homogeneous hydrogen-helium mixture is insufficient for describing Saturn's deep interior, and that the innermost region is likely composed of heavier materials, possibly in the form of a heavy element core.  
Table 3 summarizes the best fit composition value, and the composition range that includes the uncertainty factor for all the considered rotation periods and surface temperatures. \\
{\bf [Table. 3]}\\
We find that faster rotation results in lighter composition, i.e., a larger hydrogen mass fraction. Higher surface temperatures lead to 
a larger mass mixing ratio of helium. The best fit values of the hydrogen mass fraction are found to range from X=0.82 to 0.65. In principle, the average ratio of hydrogen to helium in Saturn should be similar to the abundance of the proto-Sun, X$\sim$0.725 (\citealp{Bahcall95}). Fits which produce a hydrogen mass fraction larger than solar can be excluded as being unrealistic. 
We find that several combinations of rotation period and 1 bar temperature can provide hydrogen mass fractions which are close to the proto-Sun value. In some cases, the helium mass fraction is found to be larger than the proto-Sun value. This enrichment in helium implies the existence of heavier elements in Saturn's interior. In these cases we estimate the mass of heavy elements by subtracting the helium mass based on the proto-Sun composition from the ''enriched helium'' abundance; since the mass fraction of hydrogen is smaller than solar for these models our results give only an upper bound for the mass of heavy elements. The upper bounds on the mass of heavy elements are given in Table 3. The maximum mass of heavy elements accounts for the lower bound of the hydrogen mass fraction over the entire composition range given in column 4. The total maximum mass of heavy elements is found to range from $\sim$ 6 to 20 M$_{\oplus}$.\\

In the following section we repeat the procedure described here for different pressure regions in the interior. This enables a determination of whether Saturn's composition varies with depth. We find that the helium to hydrogen ratio in Saturn's upper atmosphere can be significantly smaller than the proto-solar value possibly due to sedimentation of helium towards the center. 

\subsection{The distribution of helium}
The mass mixing ratio of helium in Saturn's atmosphere is uncertain.  Radio occultation measurements and analysis of spectra from Voyager IRIS found that helium is depleted in Saturn's atmosphere, with Y= 0.06 $\pm$ 0.05 (\citealp{Conrath84}). 
More recent calculations using Voyager data have led to higher values ranging from $Y=0.18$ to $0.25$ (\citealp{Conrath00}). These values are still lower than the protosolar value (Y$_{proto}\sim$ 0.275). 
It has been suggested by several authors (\citealp{Stevenson75}; \citealp{Stevenson77b,Stevenson77a}; \citealp{Fortney03,Fortney04}) that the depletion of helium in Saturn's atmosphere is a consequence of helium separation from hydrogen. If helium becomes insoluble in hydrogen, it can coagulate to form helium droplets that settle towards the planet's center (due to larger density). Helium separation provides an explanation for the low helium abundance in Saturn's atmosphere, and it also offers an additional energy source that seems necessary to explain the long-term evolution of Saturn (\citealp{Fortney03}). It is therefore possible that the average ratio of hydrogen to helium in Saturn is similar to the proto-Sun value, but that the distribution of helium is not uniform throughout the interior. \\
In this section we investigate whether the empirical EOS suggests a dependence of Saturn's composition on depth in its interior. We focus on three different pressure regions: the first represents the uppermost atmosphere and is defined to be from $logP(Mbar)=-6$ to $logP(Mbar)=-4$. In this pressure region hydrogen is in the molecular form. 
The second pressure region covers most of the planetary mass (99\%) ranging from $logP(Mbar)=-3$ to $logP(Mbar)=1.12$. This pressure region excludes the very low pressure region in which our interior empirical model is most uncertain. The last region is the innermost part of the planet, ranging from $logP(Mbar)=0$ to the center of the planet, which is found to be at $logP(Mbar)\sim1.12$ (the exact value depends on the assumed rotation rate). The transition from molecular hydrogen to metallic hydrogen occurs at pressure of $\sim 1$ Mbar, and it has been suggested that in the metallic region helium is most insoluble (\citealp{Hubbard85}; \citealp{Stevenson82}).\\
Again, we consider surface temperatures that range from 130 to 145 K at 1 bar, and rotation periods between 10h:32m:35s and 10h:41m:35s. We find that the upper envelope is depleted in helium and that the helium mass fraction increases significantly with depth. The best fit helium mass fraction in the low pressure region is found to range from 0 to 13\% in agreement with observations (Conrath et al., 1984). In the pressure region that covers most of the planetary mass the helium mass fraction values are found to be larger than the ones found in the previous section when the entire pressure range of Saturn was considered. This suggests that the composition of Saturn is inhomogeneous, with the mass of helium or other heavy elements increasing with depth. In the high pressure region, more than 70\% of the material is helium. However, since only hydrogen and helium are considered, this  large helium concentration probably represents an enrichment of heavier elements. The presence of a solid core would reduce the helium mass fraction to lower values than found here. However, our results still suggest helium depletion in the upper atmosphere regardless of the composition of the deep interior (see section 5 for further discussion). Table 4 presents the compositions and their uncertainty factors for all the considered cases. \\
While the bulk composition (see Table 3) varies significantly with the assumed surface temperature, we find that the composition is insensitive to the assumed surface temperature in the high pressure regions. 

{\bf [Table. 4]}\\

\section{Precession of Saturn's Pole}
Saturn's axis is tilted to its orbital plane.  The solar torque exerted on Saturn's oblate figure and on its equatorial satellites results in a precession of the planet's axis of rotation. Since the orbit plane of Saturn is not fixed in space, the precession of Saturn's pole is not constant, but changes periodically, slowly over time (\citealp{Ham04b}). The precession rate of a planet depends on both its moment of inertia and its rotation rate (and the torques driving the precession of the rotation axis). For Saturn, both quantities are a priori unknown. Saturn's precession rate can be determined in different ways and has been computed by several authors. \citeauthor{Bosh94} (1994) obtained the precession rate by a combination of Voyager occultation data and ground-based stellar occultation (28 Sgr), suggesting a rate of $-0.41''$yr$^{-1}$. Three years later Bosh obtained a precession rate of $-0.52''$yr$^{-1}$ when combining the pole position known at that time (1994) with ring plane crossings (\citealp{Bosh97}). \citeauthor{Nicholson97} (1997) and \citeauthor{Nicholson99} (1999) have computed the precession rate from 22 reported times of ring plane crossings, reporting a value of $-0.51''$yr$^{-1}$. These values reflect the precession rate at the time of the observations, not the long term average. The slow variations in Titan's inclination change the torque exerted on Saturn, with a period of about $700$ years (Nicholson et al., 1999).  As a result, the measurements must be extrapolated with a model to estimate the long term average of Saturn's precession. Currently,  the torque seems to be at its minimum value, resulting in a minimum in the rate of Saturn's pole precession,  $\sim$ 68$\%$ of the long-term (secular) value (\citealp{Vienne92}, 1992; Nicholson et al., 1999). \\
In this section we apply the moment of inertia derived from the interior models, for each assumed rotation rate, to derive the long-term precession rate of Saturn's pole. The predicted precession period of Saturn's pole due to the solar torque acting on the angular momentum of the Saturnian system is $\sim 1.76 \times 10^6$ years (French et al. 1993). Recently, \citeauthor{Jac07} (2007) has computed the precession rate from the rigid body rotational equations of motion, 
including the torques from the Sun, Titan and Iapetus. Jacobson obtained a long term (average value) 
of $-0.732''$yr$^{-1}$, suggesting a precession period of $\sim 1.77 \times 10^6$ years.\\
Following the equations in \citeauthor{French93} (1993) we define the precession period by (\citealp{Ward75}):
\begin{equation}
P = \frac{ 4 \pi \gamma' \omega}{3 J_2' n_s^2 cos\epsilon} 
\end{equation}
where $\gamma$ is the planet's moment of inertia, $\omega$ is Saturn's angular velocity, $n_S$ is Saturn's mean motion and $\epsilon$ is its obliquity.
To include the influence of the equatorial satellites, an effective second gravitational moment, $J_2'$, and an effective moment of inertia, $\gamma'$, of the Saturian system are used:
\begin{equation}
J_2' = J_2 + \frac{1}{2} \sum_i \frac{m_ja_j^2}{M_sR_s^2} 
\end{equation}
where $M_s$ is Saturn's mass, $R_s$ its equatorial radius, and $m_j$, $a_j$ are the mass and semimajor axis of the j'th satellite, respectively. 
The effective moment of inertia is given by,
\begin{equation}
\gamma' = \gamma + \sum_i \frac{m_ja_j^2 n_j}{M_sR_s^2 \omega} 
\end{equation}
where $n_j$ is the satellite's mean motion.
The physical parameters used in our computation are summarized in Table 5. Saturn's equatorial satellites' masses and radii are given in Table 6.  

{\bf [Table. 5]}\\
{\bf [Table. 6]}\\

To find the precession rate we use the normalized moment of inertia $\gamma$ obtained from the empirical interior models of section 2. Table 7 presents the computed precession rate of Saturn for different values of $\gamma$ and rotation period. As previously, the rotation periods range from 10h:32m:35s to 10h:41m:35s. Figure 4. shows the normalized moment of inertia and the calculated precession rate values as a function of rotation period. 

{\bf [Table. 7]}\\
{\bf [Fig. 4]}\\

The calculated values can be compared to different long-term precession rates available in literature: $-0.7427''$yr$^{-1}$ (French et al., 1993, after modifying the obliquity to the value presented in Table 1), $-0.75''$yr$^{-1}$ (Ward \& Hamilton, 2004) and $-0.732''$yr$^{-1}$ (Jacobson, 2007). We find that a precession rate of about $-0.754''$yr$^{-1}$ is predicted for self-consistent values of rotation rate and derived moment of inertia for all the rotation periods considered here. Our models suggest that the long-term value of Saturn's pole precession period is $\sim 1.72 \times 10^6$ years.

\subsection{Discussion and Conclusions}
We present new models of Saturn's interior with rotation periods between 10h:32m:35s and 10h:41m:35s. The models are derived using the 'theory of figures' (Zharkov \& Trubitsyn, 1978) with density profiles that are represented by a 6th degree polynomial. The interior models fit both the measured Saturnian gravitational field and the atmospheric model of Lodders \& Fegley (1998), providing an empirical polynomial density distribution of Saturn's interior (empirical EOS).\\
Using an EOS of a hydrogen and helium mixture, we find the hydrogen-helium mixing ratio that can best match the empirical $\rho(p)$ relation. Due to uncertainties in the best fit composition value and the empirical EOS we present a range of the 'best fit composition'. Since only hydrogen and helium are considered, high-Z material is  effectively included in the helium mass fraction. When the helium mass fraction exceeds the proto-Sun value we evaluate the mass of heavy elements in the interior. The global hydrogen mass fraction is found for surface temperatures that range from 130 to 145 K, and rotation periods between 10h:32m:35s and 10h:41m:35s.  
We find that the maximum mass of heavy elements ranges from $\sim$ 6 to 20 M$_{\oplus}$.\\
We look for the 'best fit composition' in different pressure regions. Helium is found to be depleted in the upper envelope, in agreement with observations and theoretical models (\citealp{Conrath00, Saumon04, Guillot05}). Higher pressure regions require larger helium mass fractions. In the deep interior, the helium mass fraction was found to increase considerably, suggesting that this region is significantly enriched with heavier elements, possibly in the form of a heavy element core. The depletion of helium in the upper atmosphere supports the idea of helium differentiation from molecular hydrogen in Saturn's envelope (Stevenson, 1975). This process not only explains measurements but it also provides an additional energy source insuring that Saturn's evolution is consistent with the age of the solar system (\citealp{Fortney03}).\\
 
Several simplifications have been made in this work. First, the empirical density distribution is given by a 6th degree polynomial in fractional radius. As a result, discontinuities in density are smoothed out (see Anderson \& Schubert, 2007 for further details) and the density in the deep interior might be underestimated, especially in the core region. Interior models that include discontinuities (a core) would require a smaller enhancement of helium (and any other high-Z material) in the deep interior. However, helium would still be expected to be depleted in the outer region and present in larger concentrations in the deep interior (\citealp{Fortney03}).
In addition, it would be desirable to include the high-Z material when calculating the 'best fit composition' so the distribution of heavy elements within the planet could be better estimated. The presence of a heavy element core can then be included as well. \\ 
The estimated hydrogen to helium mixing ratio was found using an adiabatic hydrogen and helium EOS based on the calculations of Saumon et al. (1995). Describing the interior by an adiabat is valid as long as the planet is fully convective.  Convection results in a small superadiabaticity, so the specific entropy is expected to be constant throughout the entire planet (\citealp{Hubbard73,Saumon04}). However, convection might be suppressed in certain regions due, for example, to the magnetic field, rotation, and compositional gradients (\citealp{Guillot04,Saumon04}). In the equation of state used in this work, the transition from molecular to metallic hydrogen is assumed to be continuous (Saumon et al., 1995). Thus, if the transition is first order, a discontinuity in entropy can occur, and the assumption of a fully isentropic planet breaks down (\citealp{Fortney06}). Convection can also become inefficient as a result of helium differentiation. In this case the entropy of the upper atmosphere will no longer represent the entropy of the deep interior, and a two-layer model would be required (\citealp{Fortney03}). Finally, different equations of state predict densities that can vary by up to $20\%$ at temperatures and pressures relevant to the planetary interior (\citealp{Saumon04, Guillot04,Militzer06}). \\
Improved modeling would be possible when updated atmospheric data for Saturn become available. Updated data from Cassini could lead to a smaller error in the atmospheric density resulting in a more accurate atmospheric model. In a similar way, a better determination of the temperature at the 1 bar level would provide a stronger constraint for the possible adiabats, and therefore, interior models. \\
We compute the (long-term) precession rate of Saturn's pole based on the moment of inertia values from the empirical interior models. The computed normalized moment of inertia varies with the assumed rotation period. We consider rotation periods ranging from 10h:32m:35s to 10h:41m:35s. For the four assumed rotation periods we get a precession rate value of  $\sim-0.754''$yr$^{-1}$, decreasing slowly with rotation period, suggesting that the precession period of Saturn's pole is about $\sim 1.72 \times 10^6$ years. 

\newpage


\newpage

\begin{table}[h!]
\begin{center}

{\renewcommand{\arraystretch}{0.8}
\vskip 8pt
\begin{tabular}{l c c c c}
Saturn Rotation Period & 10h:32m:35s & 10h:35m:35s & 10h:38m:35s & 10h:41m:35s
\\[1pt]
\hline
$a$ (km) & 60356.2 & 60304.3 & 60253.4 & 60203.5
\\[1pt]
$R$ (km) & 58255.4 & 58223.4 & 58191.9 & 58161.1
\\[1pt]
$q$ & 0.158851 & 0.156949 & 0.155085 & 0.153257
\\[1pt]
$m$ & 0.142835 & 0.141256 & 0.139706 & 0.138182 
\\[1pt]
$\rho_0$ (kg m$^{-3}$) & 686.276 & 687.409 & 688.525 & 689.621
\\[1pt]
$p_0$ (Mbar) & 4.46847 & 4.47832 & 4.48800 & 4.49754
\\[1pt]
$J_2$ (10$^{-6}$) & 16276.6 & 16304.6 & 16332.2 & 16359.2
\\[1pt]
$J_4$ (10$^{-6}$) & -934.2& -937.4 & -940.6 & -943.7
\\[1pt]
$J_6$ (10$^{-6}$) & 85.9 & 86.4 & 86.8 & 87.2
\\[1pt]
\hline
\end{tabular} 
}

\caption{\label{geoid} 
Fifth order (in smallness parameter, q defined in text) reference geoid for four rotation periods that span the six-minute interval of possible periods. The polar radius is fixed at the value 54,438 km, the polar radius of the 100 mbar isosurface (\citealp{Lindal1985}). The harmonic coefficients $J_{2n}$ are obtained from the measured values which are given for a reference equatorial radius of 60,330 km (Jacobson et al.~2006), according to $a^{2n}J_{2n}$=(60,330 km)$^{2n}J_{2n}$(measured). Since the equatorial radius, $a$ of the reference geoid depends on the rotation period, so do the values of $J_{2n}$. The equatorial radius, $a$ and the values of $J_{2n}$ are determined iteratively using the above relation and a second equation derived from equation (1) evaluated at the pole and the equator. The values of $J_{2n}$ from Jacobson et al.~(2006) in units of $10^{-6}$ are $J_2=16290.71\pm0.27$, $J_4=-935.8\pm2.8$, $J_6=86.1\pm9.6$.  The parameter $p_0$ is a characteristic pressure defined in the text.     
}

\end{center} 
\end{table}

\newpage
\begin{table}[h!]
\begin{center}
{\renewcommand{\arraystretch}{0.8}
\vskip 8pt
\begin{tabular}{l c c c c}
Saturn Rotation Period & 10h:32m:35s & 10h:35m:35s & 10h:38m:35s & 10h:41m:35s
\\[1pt]
\hline
$k_0$ & 6.459802 & 6.583734 & 6.712154 & 6.845172
\\[1pt]
$k_2$ & -35.442449 & -38.344462 & -41.325925 & -44.380760
\\[1pt]
$k_3$ & 12.828048 & 18.625924 & 24.583873 & 30.657281
\\[1pt]
$k_4$ & 110.378345 & 110.132162 & 109.835175 & 109.590534 
\\[1pt]
$k_5$ & -156.131034 & -162.169863 & -168.296182 & -174.607977
\\[1pt]
$\Delta J_2$ (10$^{-6}$) & 0.05 & 0.06 & 0.06 & 0.07
\\[1pt]
$\Delta J_4$ (10$^{-6}$) & 0.88 & 0.95 & 1.02 & 1.09
\\[1pt]
$\Delta J_6$ (10$^{-6}$) & 3.47 & 3.75 & 4.03 & 4.32
\\[1pt]
$\gamma$ & 0.222711 & 0.223809 & 0.224902 & 0.225990
\\[1pt]
\hline
\end{tabular} 
}
\caption{\label{model} 
Best-fit values for the polynomial coefficients that describe the variation of normalized density $\eta$ as a function of normalized mean radius $\beta$. The coefficients fit a model atmosphere (\citealp{Lodders98}) between 1000 mbar and 100 mbar with an assumed error of 100\% on the atmospheric density. They also fit the gravitational harmonics $J_2$, $J_4$, and $J_6$ with an assumed error given by the covariance matrix for $J_{2n}$ (Jacobson et al.~2006). Saturn's normalized axial moment of inertia $\gamma = C/MR^2$ is computed from the density polynomial. The converged residuals for each rotation period are given by $\Delta J_{2n}$. The fit to both the atmosphere and the harmonics is best for the shortest period of 10h:32m:35s, but all fits are acceptable, within one standard error of the measured value. The residuals in the normalized atmospheric density are on the order of 10$^{-6}$ and are not shown. All four models fit the atmospheric density to well within the assumed error of 100\%. The failure to fit the harmonics exactly is a measure of the incompatibility between the model atmosphere and the gravitational field.                  
}

\end{center} 
\end{table}

\begin{table}[h!]
{\scriptsize
\centering
{\renewcommand{\arraystretch}{0.8}
\begin{tabular}{|c|c|c|c|c|} \hline
\centering
Saturn Rotation Period &  T(1 bar) [K] & Best Fit Value & Composition Range & Maximum Mass of Heavy Elements [M$_{\oplus}$]\\
   \hline
\multirow{4}{*}{10h:32m:35s} & 130 & 0.82 & 0.66-0.98 & 6.2 \\
& 135 & 0.74 & 0.55-0.93 & 16.6\\
& 140 & 0.71 & 0.53-0.89 & 18.6\\ 
& 145 & 0.67 & 0.50-0.84 & 21.4\\ 
\hline
\multirow{4}{*}{10h:35m:35s} & 130 & 0.80 & 0.65-0.95 & 7.1\\
& 135 & 0.74 & 0.57-0.91 & 14.8\\
& 140 & 0.73 & 0.60-0.86 & 12.0\\ 
& 145 & 0.67 & 0.52-0.82 & 19.5\\
\hline
\multirow{4}{*}{10h:38m:35s} & 130 & 0.76 & 0.59-0.93 & 12.9\\
& 135 & 0.74 & 0.59-0.89 & 12.9\\
& 140 & 0.71 & 0.58-0.84 & 13.8\\ 
& 145 & 0.66 & 0.52-0.80 & 19.5\\
\hline
\multirow{4}{*}{10h:41m:35s} & 130 & 0.75 & 0.58-0.92 & 13.8\\
& 135 & 0.73 & 0.59-0.87 & 12.9\\
& 140 & 0.67 & 0.52-0.82 & 19.5\\ 
& 145 & 0.65 & 0.51-0.79 & 20.5\\ 
\hline
\end{tabular}
}
\caption{{\small The best fit of the hydrogen mass fraction for different rotation rates of the planet's interior and temperatures at the 1 bar level ranging from 130 to 145 K.}}
 \label{tab:3}
 }
\end{table}

\begin{table}[h!]
\hspace{-0.3cm}
\centering
{\renewcommand{\arraystretch}{0.8}
\begin{tabular}{||c||c||c|c|c||}
\hline
 \footnotesize{Temperature (1 bar) } & \footnotesize{Rotation Period} & \footnotesize{best fit X value} & \footnotesize{best fit X value} & \footnotesize{best fit X value}\\
  & & \scriptsize{ $logP(Mbar)=-3\sim1.12$} &\scriptsize{ $logP(Mbar)=$(-6)-(-4)}&\scriptsize{ $logP(Mbar)=0\sim1.12$}\\
\hline\hline
130 K&10h:32m:35s & 0.58 $\pm$ 0.16& 1.00 $\pm$ 0.02& 0.28 $\pm$ 0.05 \\
\hline
135 K&10h:32m:35s & 0.54 $\pm$ 0.17&  0.99 $\pm$ 0.03& 0.27 $\pm$ 0.06\\
\hline
140 K & 10h:32m:35s & 0.53 $\pm$ 0.16&   0.97 $\pm$ 0.05& 0.27 $\pm$ 0.06\\
\hline
145 K &10h:32m:35s & 0.53 $\pm$ 0.14&   0.93 $\pm$ 0.05 & 0.27 $\pm$ 0.05\\
\hline
130 K&10h:35m:35s & 0.57 $\pm$ 0.14 & 1.00 $\pm$ 0.02& 0.28 $\pm$ 0.06 \\
\hline
135 K&10h:35m:35s& 0.54 $\pm$ 0.13&  1.00 $\pm$ 0.02& 0.27 $\pm$ 0.05\\
\hline
140 K & 10h:35m:35s & 0.53 $\pm$ 0.13&   0.98 $\pm$ 0.05& 0.27 $\pm$ 0.05\\
\hline
145 K &10h:35m:35s & 0.51 $\pm$ 0.13&   0.91 $\pm$ 0.05 & 0.27 $\pm$ 0.05\\
\hline
130 K&10h:38m:35s & 0.54 $\pm$ 0.14& 1.00 $\pm$ 0.02 & 0.28 $\pm$ 0.04 \\
\hline
135 K&10h:38m:35s & 0.53 $\pm$ 0.14&  1.00 $\pm$ 0.02& 0.27 $\pm$ 0.05\\
\hline
140 K & 10h:38m:35s & 0.53 $\pm$ 0.13&   0.96 $\pm$ 0.04& 0.27 $\pm$ 0.05\\
\hline
145 K &10h:38m:35s & 0.51 $\pm$ 0.15&   0.89 $\pm$ 0.04 & 0.27 $\pm$ 0.05\\
\hline
130 K&10h:41m:35s & 0.54 $\pm$ 0.14& 1.00 $\pm$ 0.03 & 0.28 $\pm$ 0.07 \\
\hline
135 K&10h:41m:35s & 0.53 $\pm$ 0.13&  1.00 $\pm$ 0.03&  0.28 $\pm$ 0.07\\
\hline
140 K & 10h:41m:35s & 0.51 $\pm$ 0.13&  0.94 $\pm$ 0.05& 0.27 $\pm$ 0.06\\
\hline
145 K &10h:41m:35s & 0.48 $\pm$ 0.13&  0.87 $\pm$ 0.04& 0.27 $\pm$ 0.06\\
\hline
\end{tabular}
}
\caption{{\small Best fit of composition for different pressure regions assuming different surface temperatures and rotation periods.}} \label{tab:3}
\end{table}

\begin{table}[h!]
\centering
{\renewcommand{\arraystretch}{0.8}
\begin{tabular}{|c|c|}
\hline
GM$_{Saturn}$ (km$^3$s$^{-2}$)&37,931,207.7$^{(1)}$\\
GM$_{Sun}$ (km$^3$s$^{-2}$)&1.3712440018$\times 10^{20}$$^{(2)}$\\
J$_2$&16290.71$^{(3)}$\\
R$_s$, Saturn equatorial radius (km)&60268$^{(2)}$\\
a$_s$, Saturn semimajor axis (km)& 1.460268$\times 10^{9}$$^{(2)}$\\
$\epsilon$, obliquity (deg)&26.73919$^{(4)}$\\
\hline
\end{tabular}
}
\caption{{\small Physical parameters: $^{(1)}$Anderson \& Schubert (2007),$^{(2)}$JPL data,$^{(3)}$Jacobson et al. (2006),$^{(4)}$Jacobson (2007)}} \label{tab:1}
\end{table}

\begin{table}[h!]
\centering
{\renewcommand{\arraystretch}{0.8}
\begin{tabular}{|c|c|c|}
\hline
$\text{Satellite}$&$\text{GM km$^3$s$^{-2}$}$&$\text{Semimajor axis ($10^3$km)}$\\
\hline
Mimas&2.530 &185.54\\
Enceladus&7.210&238.04\\
Tethys&41.210&294.67\\
Dione&73.113&377.42\\
Rhea&154.07&527.07\\
Titan&8978.19&1221.87\\
Hypreion&0.37&1500.88\\
Iapetus&120.50&3560.84\\
\hline
\end{tabular}
}
\caption{{\small Satellite data, JPL database: http://ssd.jpl.nasa.gov}} \label{tab:2}
\end{table}

\begin{table}[h!]
\centering
{\renewcommand{\arraystretch}{0.8}
\begin{tabular}{|c|c|c|} \hline
Rotation Period & $\gamma$ & Computed Precession Rate \\ \hline
10h:32m:35s & 0.222711 & $-0.7544''$yr$^{-1}$ \\ 
\hline
10h:35m:35s & 0.223809 & $-0.7543''$yr$^{-1}$\\
\hline
10h:38m:35s & 0.224902 & $-0.7541''$yr$^{-1}$\\
\hline
10h:41m:35s & 0.225990 & $-0.7540''$yr$^{-1}$ \\ 
\hline
\end{tabular}
}
\caption{{\small Precession rates for different values of normalized moment of inertia and rotation periods. }}
 \label{tab:3} 
\end{table}

\newpage
\begin{figure}
    \centering
    \includegraphics[width=6in]{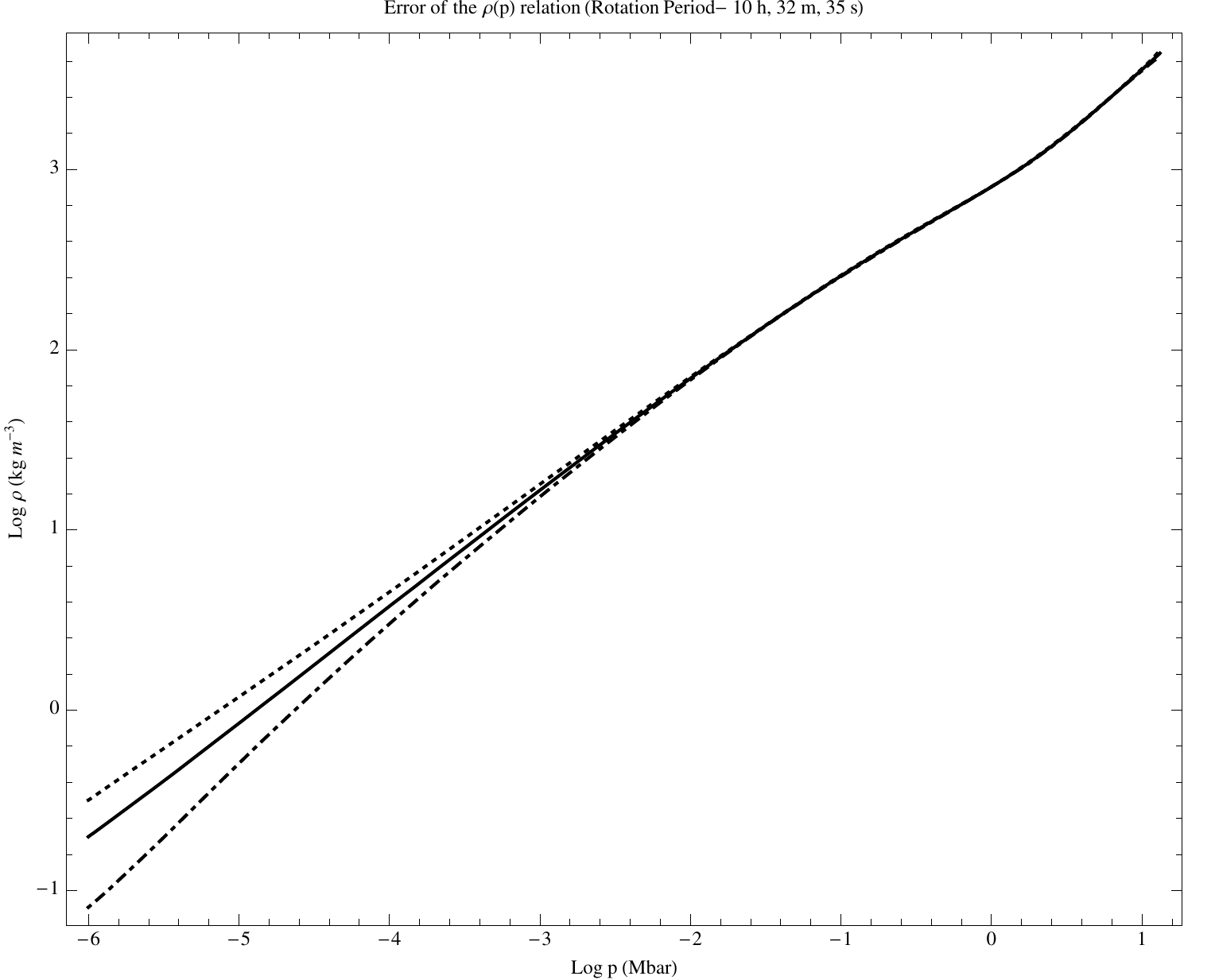}
    \caption[sat]{Saturn's empirical EOS with the error included. The solid line shows the computed $\rho(p)$, and the dotted and dashed-dotted curves present the $\rho(p)$ relation when the systematic error is added and subtracted, respectively. The area between these two curves represents the empirical EOS from the interior model. The interior model presented refers to a rotation period of 10h:32m:35s. }
\end{figure}

\begin{figure}
    \centering
    \includegraphics[width=6in]{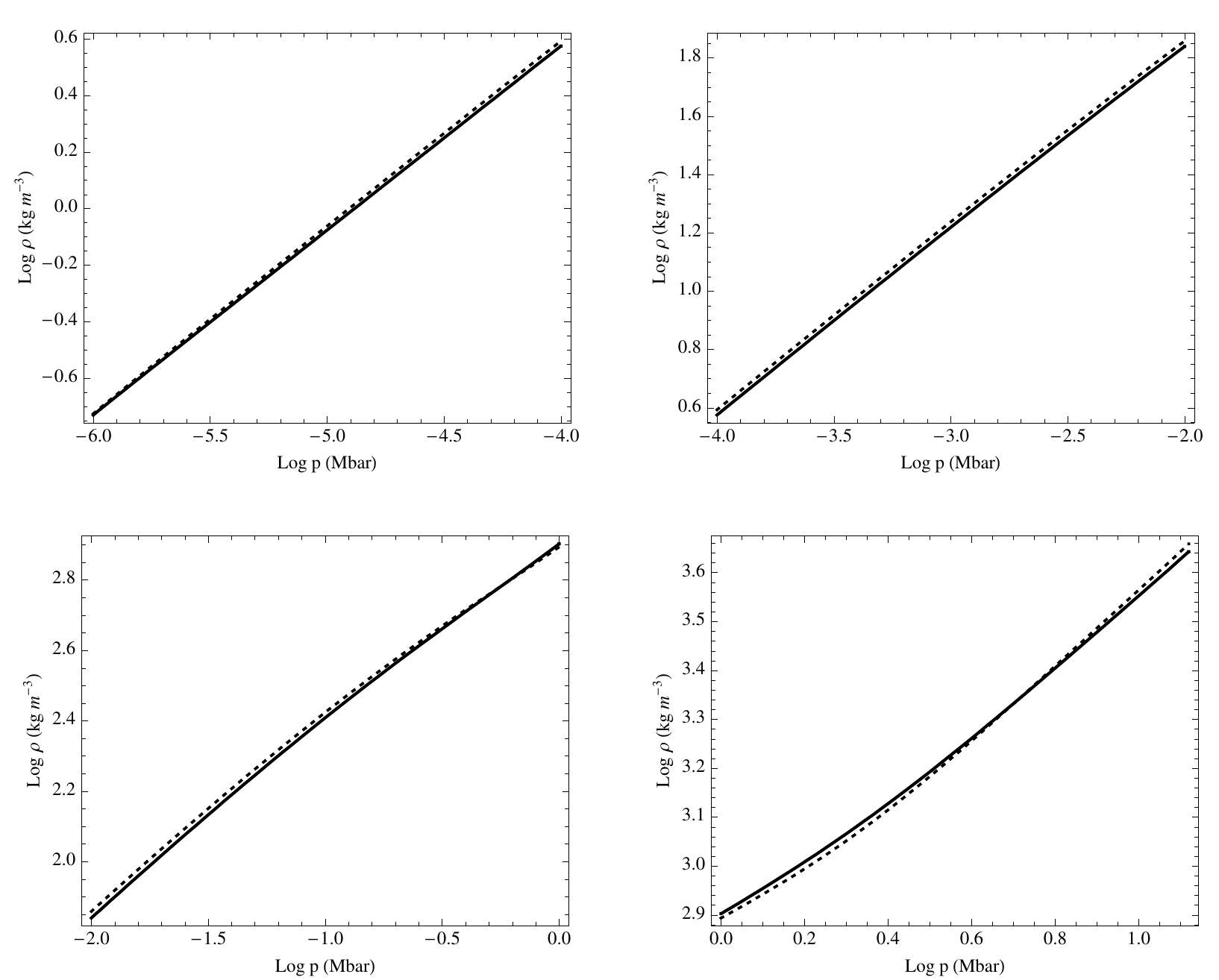}
    \caption[sat]{Saturn's empirical $\rho(p)$ relation. The solid and dotted curves represent rotation periods of 10h:41m:35s and 10h:32m:35s, respectively. }
\end{figure}

\begin{figure}
    \centering
    \includegraphics[width=6.9in ]{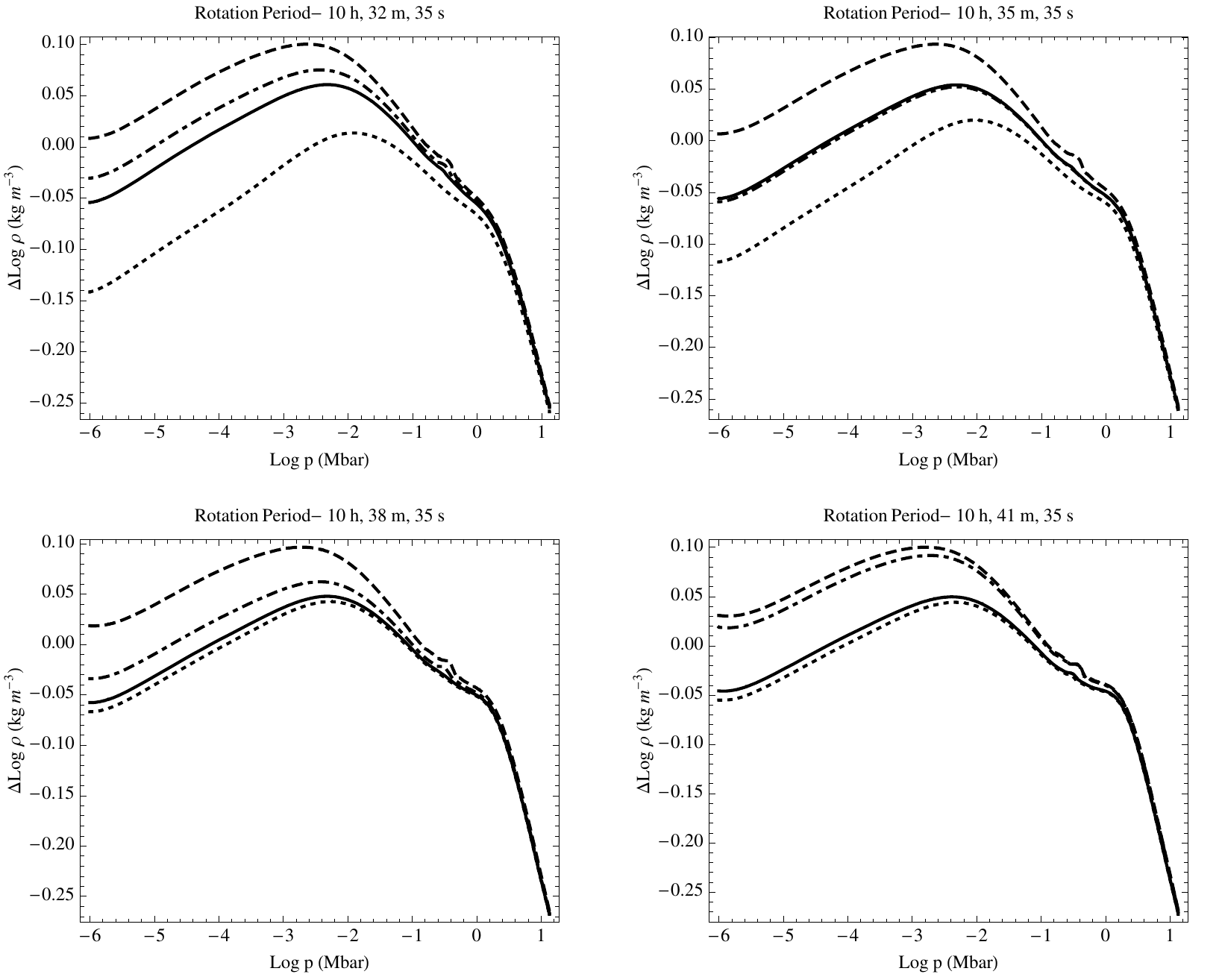}
    \caption[sat]{Saturn's empirical EOS for rotation periods ranging from 10h:32m:35s to 10h:42m:35s for different surface temperatures.
    Each plot shows the difference between the empirical EOS and the physical one, $\Delta Log\rho = (Log\rho_{H/He} - Log\rho_{empirical})$. The dotted, solid, dashed-dot and dashed curves represent temperatures of 130, 135, 140 and 145 K at the 1 bar level, respectively.}
\end{figure}

\begin{figure}
    \centering
    \includegraphics[width=6in]{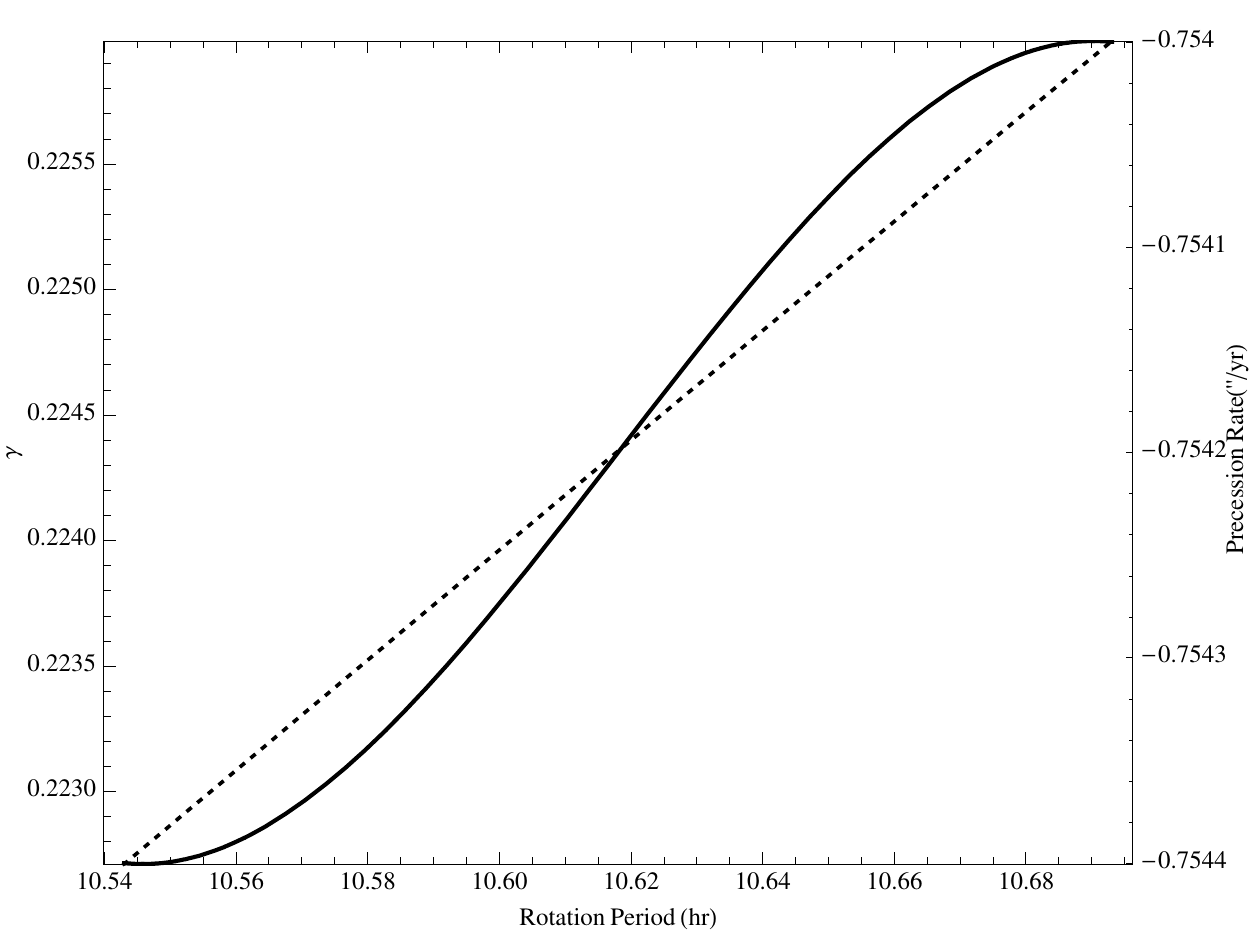}
    \caption[sat]{Saturn's normalized moment of inertia $\gamma$ (dashed) and precession rate (solid) as a function of rotation period. }
\end{figure}
\end{document}